\documentstyle [preprint,eqsecnum,aps]{revtex}
\tightenlines
\begin{document}
\draft 
%\documentstyle[pramana,floats]{ias}
%\begin{document}
%\mark{{Electronic and vibrational Raman spectroscopy of  ...}{Md. Motin Seikh, A.K. Sood and Chandrabhas Narayana}}
\title{Electronic and vibrational Raman spectroscopy of Nd$_{0.5}$Sr$_{0.5}$MnO$_3$ through the phase transitions}
\author{Md. Motin Seikh$^{\dag \ddag}$, A.K. Sood$^{\dag \#}$ and Chandrabhas Narayana$^\dag$}
\address{$^\dag$ Chemistry and Physics of materials unit, Jawaharlal Nehru Centre for Advanced Scientific Research, Jakkur P.O., Bangalore 560 064, India.\\ $^\ddag$ Solid State and Structural Chemistry Unit, Indian Institute of Science, Bangalore 560 012, India.\\ $^\#$ Department of Physics, Indian Institute of Science, Bangalore 560 012, India.}
\date{\today}
 
\maketitle

\begin{abstract}
{Raman scattering experiments have been carried out on single crystals of Nd$_{0.5}$Sr$_{0.5}$MnO$_3$ as a function of temperature in the range of 320-50 K, covering the paramagnetic insulator-ferromagnetc metal transition at 250 K and the charge-ordering antiferromagnetic transition at 150 K. The diffusive electronic Raman scattering response is seen in the paramagnetic phase which continue to exist even in the  ferromagnetic phase, eventually disappearing below 150 K. We understand the existence of diffusive response in the ferromagnetic phase to the coexistence of the different electronic phases. The frequency and linewidth of the phonons across the transitions show significant changes, which cannot be accounted for only by anharmonic interactions.}
\end{abstract}


\section{Introduction}
In the past few years the doped rare earth manganites R$_{1-x}$A$_x$MnO$_3$ (R is the trivalent rare earth and A is the divalent alkaline earth) have attracted much interest, both experimentally and theoretically, due to the strong interrelation between their structure and their electronic and magnetic properties. In this class of compounds one encounters a wealth of interrelated phenomena such as insulator-metal transitions, colossal magnetoresistance, orbital and Jahn-Teller (J-T) ordering, lattice and magnetic polarons, charge ordering and so on \cite{rao1998}.  The electrical and magnetic properties of these manganites have been understood within the framework of the double exchange (DE) of Mn$^{3+}$-O-Mn$^{4+}$ involving the hopping of the spin polarized charge between the Mn$^{3+}$ and Mn$^{4+}$ sites, which results in the ferromagnetism as well as metallic conduction \cite{zener1951}. However, DE alone is not enough to explain the metal-insulator transition in the manganites \cite{elemans1971} and mechanisms involving strong electron-phonon interactions have been proposed \cite{millis1995,millis1996,goodenough1997}. The electron-phonon interaction arises mainly  from two types of distortions of the ideal cubic perovskite structure. One is the tolerance factor, t = (r$_A$ + r$_O$)/$\sqrt2$(r$_{Mn}$ + r$_O$) (r stands for the ionic radius) originating from the cation size mismatch. The second one is the J-T  distortion of the MnO$_6$ octahedra, associated with the degenerate e$_g$ orbitals of the Mn$^{3+}$ ions. The J-T distortion moves together with the electron and a strong electron-phonon coupling tends to localize the conduction electrons resulting in a lattice polaron. There is a wealth of experimental evidence supporting the polaron formation in the perovskite manganites using various probes \cite{billinge1996,zhao1996,booth1998}, including Raman spectroscopy \cite{iliev1998}. The dynamics of this distortion is a sensitive function of the doping level and temperature. The MnO$_6$ octahedra are highly disordered in the paramagnetic insulating state and the magnitude of the distortion decreases as the temperature is reduced near the transition. The ferromagnetic metallic phase is less distorted than the paramagnetic phase. 
Furthermore, the coexistence of different electronic and magnetic phases is commonly encountered in doped rare earth manganites \cite{rao1999,woodward1999,mayr2001}. 

Raman spectroscopy has been successively employed to investigate phonons and electronic excitations in various compounds of the manganite family \cite{iliev1998,yoon1998,liu1998,dediu2000,gupta1996,liarokapis1999,gupta2002,granado2000,abrashev1999}. Iliev $\it{et}$ $\it{al}$. \cite{iliev1998} have carried out a symmetry analysis of the important modes of the parent orthorhombic LaMnO$_3$ and structurally related perovskite compounds. Raman spectra of the doped manganites show broad bands at positions close to those of the strong Raman lines for the parent RMnO$_3$ compounds with the P$\it{nma}$ space group. The broadness of the modes over a large temperature window ruled out the effect of thermal broadening and is attributed to the phonon density-of-state (DOS) related to strongly distorted oxygen sublattice. Gupta $\it{et}$ $\it{al}$.  \cite{gupta1996} have carried out Raman investigations on La$_{0.7}$Sr$_{0.3}$MnO$_3$ and observed a broad peak centered around 2100 cm$^{-1}$, which has been attributed to electronic excitation. The electronic contribution to the Raman scattering across the metal-insulator transition has also been studied  in other doped manganite systems as well \cite{yoon1998,liu1998}.  In this paper, we present the temperature-dependence of the electronic and vibrational Raman scattering in single crystals of Nd$_{0.5}$Sr$_{0.5}$MnO$_3$ which undergoes a phase transition at 250 K (T$_C$) from a paramagnetic insulating (PMI) state to a ferromagnetic metallic (FMM) state. On further cooling,  it undergoes a first order phase transition from the FMM state to a charge-ordered CE-type antiferromagnetic (AFM) insulating state at 150 K (T$_{CO}$) \cite{kuwahara1995,kuwahara1996}. Initially, the crystal structure of Nd$_{0.5}$Sr$_{0.5}$MnO$_3$  at room temperature was assigned to be the space group P$\it{mma}$, where the distortions of the ideal cubic perovskite structure correspond to the rotation of the octahedra with respect to the [101] and [010] directions \cite{kuwahara1995,kuwahara1996}. Subsequent studies showed that the space group should be I$\it{mma}$, where the rotation of the octahedral takes place along [101] direction \cite{rao1999,woodward1999,ritter2000}. As the temperature is lowered below 250 K, there is an increase in the {\it a} and {\it c} lattice parameters with a simultaneous decrease in the {\it b} parameter, without any change in symmetry \cite{woodward1999}. On cooling below the T$_{CO}$, Nd$_{0.5}$Sr$_{0.5}$MnO$_3$ shows a structural transition to a monoclinic structure P$\it{2_1/m}$.  A comprehensive Brillouin scattering temperature-dependent study \cite{seikh2003} of the surface Rayleigh wave frequency shows an anomaly at 200 K, which is attributed to the appearance of an intermediate A-type antiferromagnetic phase before the charge-ordering transition. Our present Raman results show signatures of both the transitions at 250 K and 150 K. Raman spectra reflect in a unique way the structural disorder is induced by the non-coherent J-T distortion and its variation with temperature. The present study also provides evidence for the coexistence of small and large polarons in the FMM phase of Nd$_{0.5}$Sr$_{0.5}$MnO$_3$. Below the charge-ordering transition new phonon modes are seen.

\section{Experimental details}
Polycrystalline powders of Nd$_{0.5}$Sr$_{0.5}$MnO$_3$ were prepared by the solid-state reaction of stoichiometric amounts of neodymium acetate, strontium carbonate and manganese dioxide. The initial materials were grounded and heated at 1000 $^\circ$C with two intermediate grindings for 60 hrs. This is followed by a heating at 1200 $^\circ$C for 48 hrs. The x-ray powder diffraction pattern showed the powder to be of single phase. These polycrystalline powders were filled in a latex tube and compacted using a hydrostatic press at a pressure of 5 tons. The rods thus obtained were sintered at 1400 $^\circ$C for 24 hrs. Single crystals of  Nd$_{0.5}$Sr$_{0.5}$MnO$_3$ were grown from these rods using the floating zone melting technique, which employs SC-M35HD double reflector infrared image furnace (Nichiden Machinery Ltd., Japan). The growth rate and rotation speeds used for the crystal growth were 10 mm/hrs and 30 rpm, respectively.  The system was subjected to 2 l/min flow of air. The resistivity and magnetization measurements showed the expected transitions as reported in the literature \cite{kuwahara1995}.

Raman measurements were performed in 90$^\circ$ scattering geometry using a Jobin Yvon TRIAX 550 triple grating spectrometer equipped with a superNotch filter (Kaiser Optical Systems, Inc.) and a liquid nitrogen cooled CCD detector. The scattered light  was collected using a microscope objective (10X/0.25) and a fibre optic cable \cite{kavitha2003}. The excitation source used is a diode-pumped frequency doubled solid state Nd:YAG laser of 532 nm (Model DPSS 532-400, Coherent Inc. USA). The crystal was mounted on the cold finger end of a closed-cycle He cryostat (CTI cryogenics, USA). The sample temperature was measured with an accuracy of $\pm$ 1 K. The incident laser power was kept below 20 mW focused to a diameter of $\sim$ 50 $\mu$m. The laser heating of the sample estimated from the ratio of Stokes and anti-Stokes Raman signals is $\sim$ 20 K and has been taken into account while plotting the data. Raman spectra were collected as a function of temperature in the 320-50 K range in the cooling cycle. 

\section{Results and discussion}

Figure 1 shows the Raman spectra of Nd$_{0.5}$Sr$_{0.5}$MnO$_3$ at a few typical temperatures wherein the reduced intensity Im $\chi(\omega)$ (= observed intensity/(n($\omega$)+ 1), n($\omega$) is the Bose-Einstein factor) is plotted. It is seen that above T$_{CO}$, Raman modes associated with the optical phonons lie on a broad background, which has been attributed to collision-dominated \cite{zawadowski1990} electronic Raman scattering \cite{yoon1998,liu1998}. The spectra are fitted to 

\begin{equation}
\mbox{Im}\;\chi(\omega) = \mbox{I}_{el}(\omega) + \mbox{I}_{ph}(\omega) = \frac{A\omega\Gamma_{el}}{\omega^2 + \Gamma_{el}^2} + \sum^{n}_{i=1}(2B_i/\pi)\frac{\Gamma_i}{(\omega - \omega_i)^2 + \Gamma_i^2}
\end{equation}

\noindent
The first term arises from collision-dominated electronic scattering and the second term represents the vibrational modes. In Fig. 1, the dash-dot line represents the electronic Raman scattering, I$_{el}(\omega)$, whereas the phonon contributions are shown by the dotted line. The smooth solid line is the resultant fit including both the contributions, as in Eq. (1), for panels (a) and (b). For panels (c) and (d), there is no electronic background and hence the spectra are fitted only to a sum of Lorentzians representing optical phonons, the second term in Eq. (1). We find that above T$_{CO}$, we have to include four phonon modes (n = 4) to fit the data, whereas between T$_{CO}$ and 80 K, the spectra have to be fitted with n = 5. For the spectra at 70 and 50 K, we have to include another phonon mode at 660 cm$^{-1}$ to get a good fit of the data and the sum of Lorentzian functions.  

In our experiments, the electronic background is present only above T$_{CO}$ and it can be fitted to $I_{el}(\omega)$ as given in Eq. (1) in the entire temperature range of 320 to 150 K, where A is the symmetry-dependent amplitude and $\Gamma_{el}$ is the carrier scattering rate arising from spin fluctuations and local lattice distortions. Figure 2 shows the plots of A and $\Gamma_{el}$ against temperature.  Ideally, we should normalize A with respect to the intensity of a phonon, which does not vary with temperature to take into account any artifacts due to changes in scattering conditions.  This was not possible in the present case, since all the phonon modes seem to be changing with temperature.  Instead, we used another sample of Si crystal, kept next to the crystal of Nd$_{0.5}$Sr$_{0.5}$MnO$_3$, whose Raman bands was used as external calibration for the spectra being discussed here.  As seen in Fig. 2, the value of $\Gamma_{el}$ obtained from fit at 320 K is $\sim$120 cm$^{-1}$, similar to the reported value in Pr$_{0.63}$Sr$_{0.37}$MnO$_3$ \cite{liu1998}.  In the Raman studies of Pr$_{0.63}$Sr$_{0.37}$MnO$_3$ \cite{liu1998} and in Nd$_{0.7}$Sr$_{0.3}$MnO$_3$ \cite{choi2003}, it is found that the  electronic Raman background becomes flat below T$_C$. This flat background is given by the same function I$_{el}(\omega)$ as in Eq. (1), but with frequency dependent $\Gamma_{el} = \Gamma_0 + \alpha \omega^2$, where the parameter $\alpha$ reflects electron correlation effects \cite{liu1998}. Our experiments, however, show a different behavior, wherein the electronic background below T$_C$ in the FMM phase is also given by $I_{el}(\omega)$ as in Eq. (1), with a frequency-independent $\Gamma_{el}$. The fact that the diffusive electronic background in the metallic state is similar to that in  the paramagnetic insulating phase implies that even below T$_C$, the charge carries are not completely delocalized and that there is substantial scattering of carriers from local lattice distortions and spin fluctuations. This may be due to the coexistence of the FMM, PMI and AFM phases \cite{woodward1999,ritter2000}. The amplitude parameter A and scattering rate $\Gamma$ decrease gradually with lowering of temperature (Fig. 2), showing that the scattering of carriers reduces gradually below T$_C$.

The main features observed in the spectrum at room temperature are the four bands centered around 200 cm$^{-1}$ (R$_1$), 340 cm$^{-1}$ (R$_2$), 450 cm$^{-1}$ (R$_3$) and 600 cm$^{-1}$ (R$_4$). These broad bands are seen in the insulating high-temperature phases of all the disordered manganites, independent of the crystallographic symmetry (for example, orthorhombic \cite{liarokapis1999,granado2000,choi2003} or rhombohedral \cite{gupta2002,abrashev1999} phases), the level of doping (undoped \cite{liarokapis1999,granado2000}, doped \cite{dediu2000,gupta2002}) or on the element at the A site.  The assignment of these modes is based on lattice dynamic calculations \cite{iliev1998} and similar to Choi et al \cite{choi2003}. The Raman modes are related to the motions of the MnO$_6$ octahedra. The R$_1$ mode is associated with the in-phase rotational mode of the MnO$_6$ octahedra, the R$_2$ mode with the octahedral tilt of the MnO$_6$ octahedra, the R$_3$  mode with the out-of-phase bending of the MnO$_6$ octahedra and the R$_4$ mode with the symmetric stretching of the basal oxygen ions.

The Raman spectra recorded by us can be fitted satisfactorily to the four phonon modes in the 320 to 150 K range. As the temperature is lowered below 150 K, the lineshape near the R$_3$ mode changes, suggesting  an appearance of a new mode at $\sim$ 490 K (labeled as R$_{CO1}$). Below 70 K, another band appears around 660 cm$^{-1}$ (R$_{CO2}$). Similar bands near 490 and 660 cm$^{-1}$ are also observed in the paramagnetic phase of Nd$_{0.7}$Sr$_{0.3}$MnO$_3$. The mode near 490 cm$^{-1}$ is attributed to the out-of-phase stretching mode of the Jahn-Teller distorted MnO$_6$ octahedron \cite{choi2003}. The 660 cm$^{-1}$ mode is due to the breathing mode of the octahedron. The intensities of the 490 and 660 cm$^{-1}$ bands are related to the magnitude of the static Jahn-Teller distortions \cite{carron2002} and hence their appearance below T$_{CO}$ clearly indicates the presence of Jahn-Teller distortions in the charge and orbitally ordered phase.  Choi et al \cite{choi2003} have attributed the appearance of new modes at 490 and 612 cm$^{-1}$ below T$_{CO}$ to the change of crystal symmetry from P{\it mma} to P{\it nma} and the modes at $ \sim 660$ cm$^{-1}$ to the concominecent charge and orbital ordering.

Fig. 3 shows the temperature dependence of the frequencies and the full-width at half-maxima (FWHM) of all the modes observed in the spectra. The linewidths of all the modes are rather high, possibly due to the lattice disorder and temperature-dependent dynamics. The frequency of the R$_1$ mode shows a substantial increase near T$_C$ and a gradual decrease in the FWHM across the PMI-FMM transition as shown in Figs. 3 (a) and (b). Figures 3(c) and (d) show the variations in the frequency and FWHM of the R$_2$ mode. The linewidth of the R$2$ mode shows an anomalous increase with the decrease in temperature down to T$_{CO}$, below which it decreases. A similar anomaly occurs in the FWHM of the R$_3$ mode as can be seen from Fig. 3 (f). The open symbols in Fig. 3(e), (f), (g) and (h) are associated with the R$_{CO1}$ and R$_{CO2}$ modes, respectively. The inset in Fig. 3(f) shows the temperature variation of the intensity ratio of the R$_{CO1}$ with respect to that of the R$_3$ mode, which reveals that the intensity of the R$_{CO1}$ mode increases with the lowering of temperature below T$_{CO}$. The noteworthy features in these results are:
(1) The changes in the frequencies of all the modes, R$_1$ to R$_4$, are much larger than what can be expected from anharmonic effects. (2) The changes in the linewidths of the R$_3$ and R$_4$ modes are also large. (3) The large changes in $\omega$ and $\Gamma$ with temperature reflect that electron-phonon interactions and spin-phonon interactions are large. A quantitative interpretation of the temperature dependence of the frequency and linewidths of the Raman modes is difficult at this stage. 

It is pertinent to recall the mechanism suggested to explain the anomalous temperature dependence of the Raman linewidths in charge-ordered Pr$_{0.63}$Ca$_{0.37}$MnO$_3$ \cite{gupta2002}. It involves a strong coupling between the Raman modes and spin-excitations. This coupling arises because the tilt of the neighboring MnO$_6$ octahedra changes the Mn-O-Mn bonds angle $\theta_{ij}$ between the nearest-neighbor Mn-O bonds involving Mn ions $i$ and $j$. Since the oxygen mediated overlap t$_{ij}$ between Mn-{\it d} orbitals goes as cos$\theta_{ij}$, the spin exchange constant J$_{ij}$ gets modulated. Physically, this accounts to the decay of the phonons into another phonon and two spin excitations. At temperatures higher than T$_C$, there can be short range spin correlations and the ferromagnetic clusters can also exist. As the temperature is lowered, the spin excitations are less damped and contribute to the self-energy of the phonon via the above mechanism, leading to an increase in the phonon linewidth. This will be applicable till T$_{CO}$, below which the spin excitations are modified. The reason that the temperature dependence of the linewidths of R$_1$ and R$_4$ are different from those of R$_2$ and R$_3$ modes may be because the modes R$_1$ and R$_4$ do not involve the modulation of the Mn-O-Mn bond angle and hence do not have significant spin-phonon coupling.

We will now point out the salient features of our study visa-vis that of Choi et al \cite{choi2003} on the same system. (1) In our study, electronic Raman scattering  has been quantitatively analysed and it is shown that frequency-independent $\Gamma$ in I$_{el}(\omega)$ (Eq. (1)) is applicable in PMI as well as in FMM state.  This is first such observation in the FMM state. (2)  Raman spectra were fitted by a sum of Lorentzians to get the temperature-dependence of phonon frequencies and their linewidths.  This was not done in Ref. \cite{choi2003}. We find that the phonon frequencies increase with decreasing temperature - a behavior expected from anharmonic interactions.  The change is, however, much larger than the anharmonic effects in other perovskites.  On the other hand, the lifetimes associated with these phonons show an anomalous behavior above T$_C$, except in the case of R$_1$ mode. We have given a plausible explanation of the anomalous behavior, in terms of spin-phonon interactions.

As remarked earlier, Brillouin studies reveal a transition at 200 K, which is associated with existence of two magnetic phases, namely, A-type anti-ferromagnetic phase and ferromagnetic phase.  Since Brillouin scattering is very sensitive to the magnetic order, it could be seen.  However, such is not the case in Raman scattering and hence there were no signature of 200 K transition in the Raman data.

\section{Conclusions}
A quasi-elastic Raman response is observed from 320 K to T$_{CO}$ arising from the collision-dominated scattering of the charge carriers by spin, phonons and impurities. Unlike other manganites, the nature of electronic Raman scattering is similar in the paramagnetic insulator as well as in ferromagnetic metallic phases, namely that the scattering rate $\Gamma$ is frequency-independent. This implies that the charge carriers are not fully delocalized even in the ferromagnetic metallic phase. Broad Raman bands associated with the optical phonons are seen over the electronic background. The linewidths of all the modes are very large. The temperature dependence of the modes frequencies and linewidths can not be accounted for by the anharmonic effects. The temperature-dependence  of the linewidths of two Raman modes is anomalous. We hope that our results will motivate theoretical studies to understand electron-phonon and spin-phonon interactions in manganites which exhibit both metal-insulator as well as charge-ordering transitions.

\section{Acknowledgments}
Mr. Seikh thanks Council of Scientific and Industrial Research (CSIR), Govt. of India, for a research fellowship. We thank Department of Science and Technology (DST) and BRNS (DAE), India, for financial support.

\begin{figure}
\label{fig1}
\caption{Raman spectra of Nd$_{0.5}$Sr$_{0.5}$MnO$_3$ at different temperatures. The data is corrected by Bose-Einstein thermal factor [n($\omega$)+ 1] = [1-e$^{-(\hbar\omega/k_BT)}$]$^{-1}$ to get Im $\chi$ = S$(\omega)$/[n($\omega$)+ 1]. The solid line over experimental data is the total fit. The dash dot line shows the electronic part given by I $(\omega) = \frac{A\omega\Gamma}{\omega^2 + \Gamma^2}$ and the dotted lines are phonon modes fitted with Lorentzian.}
\end{figure}

\begin{figure}
\label{fig2}
\caption{Temperature dependence of (a) the $\Gamma$, related to carrier scattering rate and (b) the symmetry dependent amplitude function, A obtained by using the collision-dominated model.}
\end{figure} 

\begin{figure}
\label{fig3}
\caption{Temperature variation of the phonon mode frequencies and FWHM: (a) and (b) for R$_1$; (c) and (d) for R$_2$; (e) and (f) for both the R$_3$ (filled symbols) and R$_{CO1}$ , the new mode (open symbols); (g) and (h) for both the R$_4$ (filled symbols) and R$_{CO2}$, the new mode (open symbols). The inset in (f) shows the intensity ratio of R$_3$ and R$_{CO1}$ as a function of temperature. Solid lines are drawn as guide to the eye.}
\end{figure}

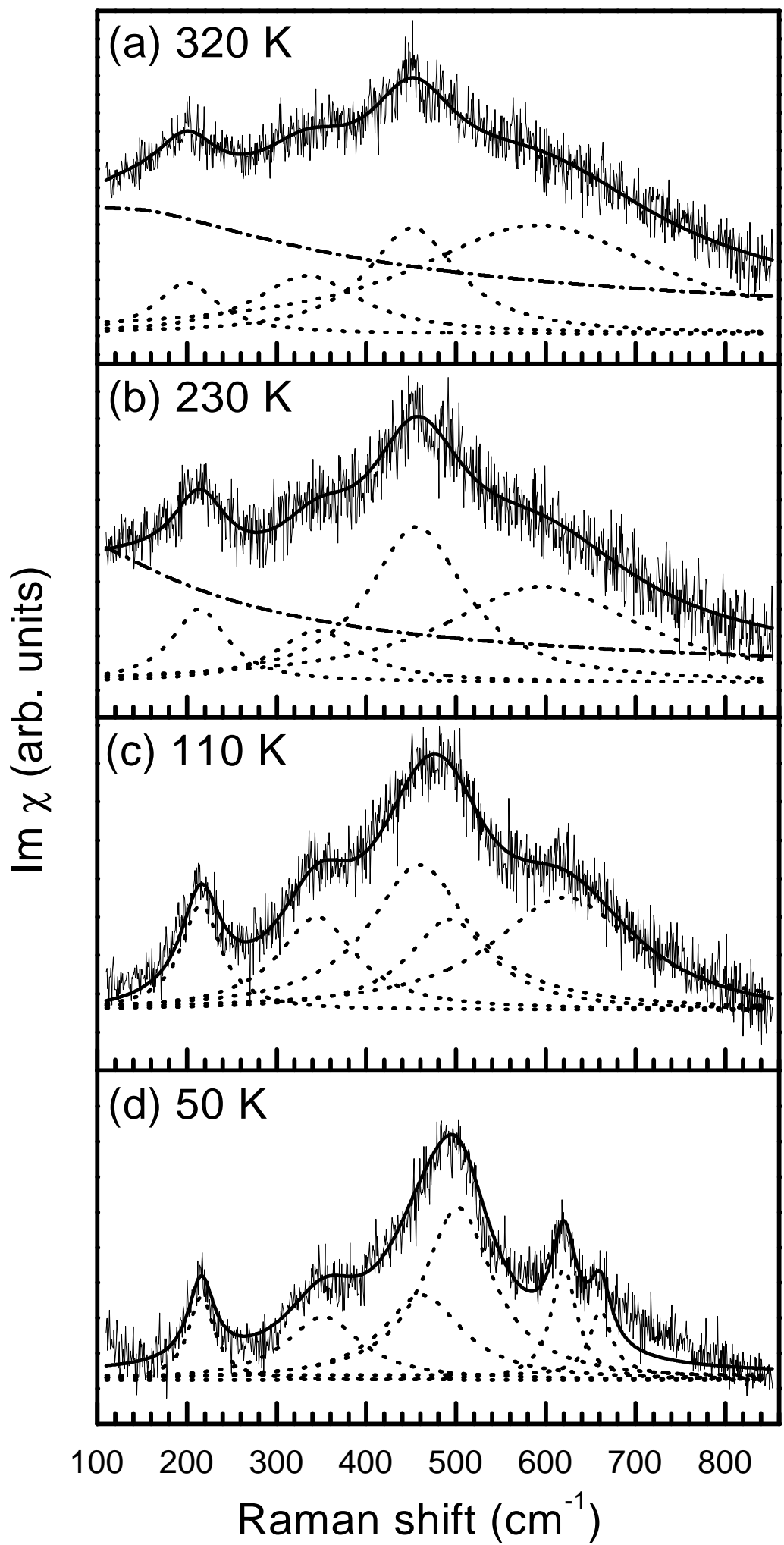
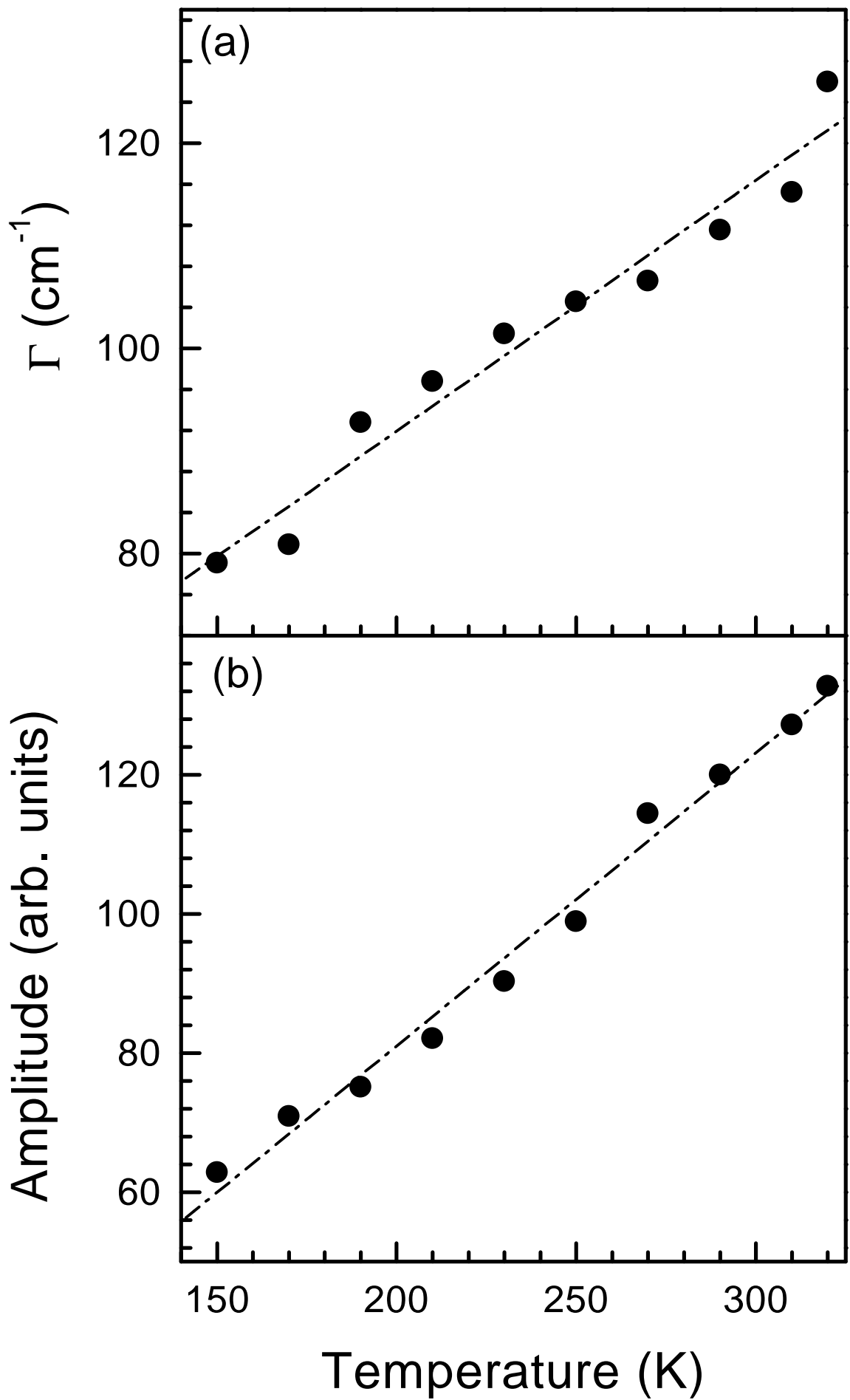
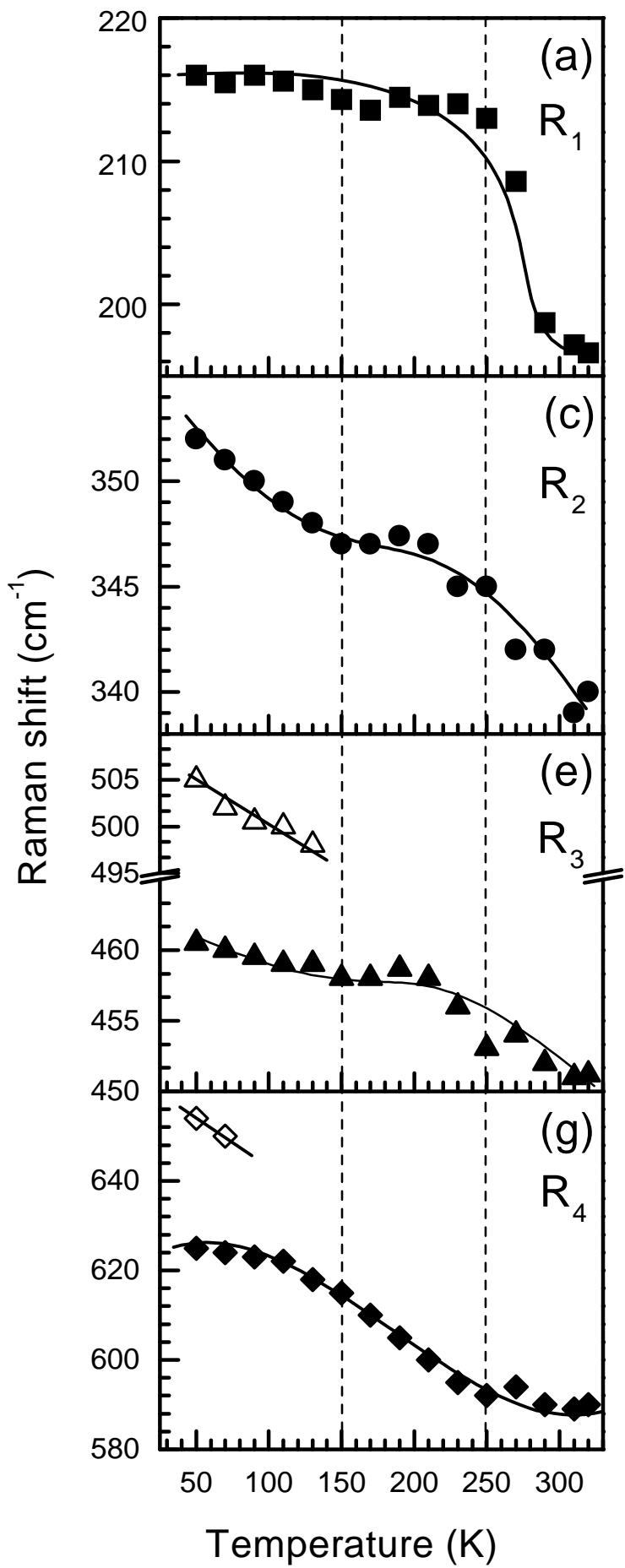
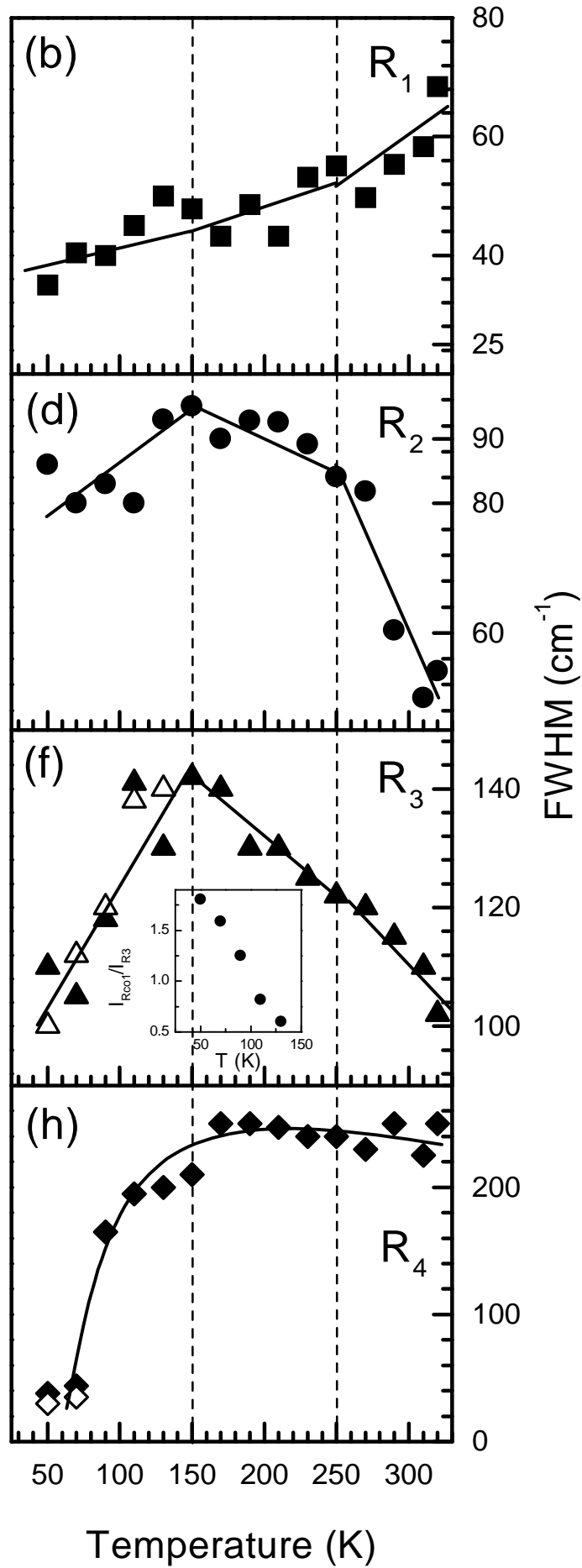
\end{document}